# Synergistic improvement of specific strength and plasticity achieved in Ti-based metallic glass designed based on quasicrystal structure*

Zhengqing Cai[1], Zijing Li[2], Shidong Feng[1, *], Limin Wang[1, 3], Riping Liu[1]

1. Center for Advanced Structural Materials, State Key Laboratory of Metastable Materials Science and Technology, Yanshan University, Qinhuangdao 066004, China
2. College of Science, Yanshan University, Qinhuangdao 066004, China
3. School of Materials Science&Engineering, Hebei University of Technology, Tianjin 300130, China

*shidongfeng@ysu.edu.cn

**Abstract**

Achieving a balance between low density, high strength, and good ductility remains a major challenge in the development of structural materials. Ti-based bulk metallic glasses (BMGs) have attracted considerable attention due to their exceptionally high specific strength. However, the intrinsic strength–plasticity trade-off has hindered their practical applications. Based on a quasicrystal-derived structural heredity and minor-element microalloying, this work realizes a synergistic enhancement of specific strength and plasticity in Ti-based BMGs. The resulting $((Ti_{40}Zr_{40}Ni_{20})_{72}Be_{28})_{97}Al_3$ BMGs demonstrate an ultrahigh specific strength of $5.34 \times 10^5$ N·m·kg$^{-1}$, establishing a new record for Ti-based BMGs, along with a plastic strain of 13%, breaking through the traditional strength–plasticity limitation of BMGs. Structural analyses show that Al microalloying effectively inherits and modulates the short-range order derived from quasicrystalline structures, thereby achieving an observed synergistic enhancement in both strength and plasticity. This work provides new insights into composition design and lightweight structural applications of Ti-based BMGs.

---

## 1 Introduction

In automotive, aerospace, and defense applications, the demand for structural materials that simultaneously exhibit low density, high strength, and high specific strength is increasingly urgent[1-3]. Bulk metallic glasses (BMGs) bypass crystalline nucleation and growth during cooling, solidifying directly into a non-equilibrium state and thus are regarded as frozen liquids. This unique disordered structure endows BMGs with strength approaching theoretical limits and high hardness, along with excellent corrosion resistance, wear resistance, and, in certain systems, favorable magnetic and electrical properties[4-6]. Over the past several decades, researchers have successfully synthesized numerous BMG systems and systematically elucidated their formation mechanisms and intrinsic properties[7-9].

Among various BMG systems, Ti-based BMGs have attracted considerable attention due to their high specific strength and good corrosion resistance, demonstrating broad application potential in critical sectors such as defense, marine engineering, and energy. However, despite achieving specific strengths up to twice that of conventional Ti alloys, existing Ti-based BMGs still exhibit limited plasticity, typically accompanied by pronounced brittle fracture. For instance, the Ti-Zr-Be-Al system currently reported to possess the highest specific strength[10] achieves a value of $4.65\times10^5$ N·m/kg but displays only ~2.1% plasticity, highlighting the inherent trade-off between strength and ductility. Consequently, achieving concurrent enhancement of both high specific strength and high plasticity in Ti-based BMGs represents a pivotal scientific challenge requiring urgent breakthroughs.

Previous studies have shown that the $(Ti_{40}Zr_{40}Ni_{20})_{72}Be_{28}$ BMG, designed based on a quasicrystal precursor, contains abundant distorted and perfect icosahedral clusters, resulting in outstanding comprehensive mechanical properties[11]. Building upon this foundation, microalloying with Al is expected to further enable synergistic optimization of properties. On one hand, Al exhibits a relatively high Poisson's ratio and has been demonstrated in Zr-Cu-Ti and Cu-based BMG systems to improve work-hardening behavior while simultaneously enhancing both strength and plasticity[12]. On the other hand, the low density of Al (2.7 g/cm$^3$) effectively reduces the overall density of the alloy, thereby enhancing its specific strength. Moreover, the large negative enthalpies of mixing between Al and Ti or Zr (-30 and -44 kJ/mol, respectively) promote the formation of stable chemical bonds, strengthening interatomic cohesion and improving resistance to deformation[13]. Additional research has also indicated that Al addition in Zr-Ti-Ni-Nb systems facilitates the formation of short-range ordered structures and modulates free volume distribution, significantly enhancing mechanical performance[14].

Based on these insights, this study employs the quasicrystal-derived $(Ti_{40}Zr_{40}Ni_{20})_{72}Be_{28}$ BMG as a foundation and introduces Al to achieve concurrent improvements in specific strength and plasticity. The goal is to transcend the current strength–plasticity performance boundary of Ti-based BMGs and provide novel strategies and theoretical foundations for the compositional design of lightweight, high-strength structural materials.

## 2 Experimental Methods

Master alloy ingots of $((Ti_{40}Zr_{40}Ni_{20})_{72}Be_{28})_{100\text{-}x}Al_x$ ($x$ = 1.5, 3, 4.5, 6, 7.5) were prepared by arc melting under a high-purity argon atmosphere (purity >99.999%). Prior to melting the master alloys, a titanium ingot was melted first to scavenge residual oxygen in the chamber. To ensure compositional homogeneity, each alloy ingot was flipped and re-melted six times. Raw materials consisted of elemental metals with purity exceeding 99.9%. Cylindrical specimens with a diameter of 2 mm were subsequently fabricated via copper mold suction casting.

As-cast amorphous specimens were sectioned using a diamond low-speed saw under water cooling. The microstructure of these as-cast alloy rods was characterized by X-ray diffractometry (XRD, Rigaku D/max-2500/PC) using Cu Kα radiation. Uniaxial compression tests were performed on an Instron-5982 universal testing machine. Compression specimens had a diameter of 2 mm and a length-to-diameter ratio of 2:1, tested at a strain rate of $5\times10^{-4}$ $s^{-1}$. To ensure data reliability, each compression test was repeated three times. Fracture surfaces and lateral morphologies of compressed specimens were examined using a Hitachi S-3400 scanning electron microscope (SEM). Additionally, 2 mm suction-cast alloy rods were cut into 0.5 mm thick circular discs under water cooling, ultrasonically cleaned multiple times with acetone and ethanol, and then subjected to differential scanning calorimetry (DSC, NETZSCH 449F3). DSC measurements used 10-20 mg samples heated at a rate of 20 K/min. For clarity, the $((Ti_{40}Zr_{40}Ni_{20})_{72}Be_{28})_{100\text{-}x}Al_x$ BMGs are hereafter denoted as Alx, according to their Al content.

## 3 Results

The microstructure and thermodynamic properties of $((Ti_{40}Zr_{40}Ni_{20})_{72}Be_{28})_{100\text{-}x}Al_x$ alloys with varying Al contents were systematically investigated. Figure 1(a) presents the X-ray diffraction (XRD) patterns of these alloys. All samples exhibit typical broad diffuse halos without any sharp Bragg peaks, confirming their fully amorphous structure. Figure 1(b) shows the differential scanning calorimetry (DSC) curves of the alloys at a heating rate of 20 K/min. Upon heating, all samples display a distinct glass transition followed by a broad endothermic peak in the supercooled liquid region. At higher temperatures, multiple exothermic peaks appear, indicating multi-stage crystallization during heating and further verifying the amorphous nature of the samples. As the Al content increases, the crystallization behavior changes markedly: at 1.5 at.% Al, two clear crystallization peaks are observed, whereas at Al contents of 3 at.% and above, the number of crystallization peaks increases to three or more, demonstrating that Al addition significantly influences the crystallization kinetics and thermal stability. In the higher-temperature region, multiple endothermic peaks corresponding to sequential melting events are also observed. Key thermophysical parameters—including glass transition temperature ($T_g$), crystallization temperature ($T_x$), supercooled liquid region width ($\Delta T_x = T_x - T_g$), melting temperature ($T_m$), and liquidus temperature ($T_l$)—were extracted from the DSC curves and are summarized in Table 1. With increasing Al atomic content, $T_g$, $T_x$, and $\Delta T_x$ all exhibit upward trends, indicating that Al incorporation effectively enhances the thermal stability of the alloy.

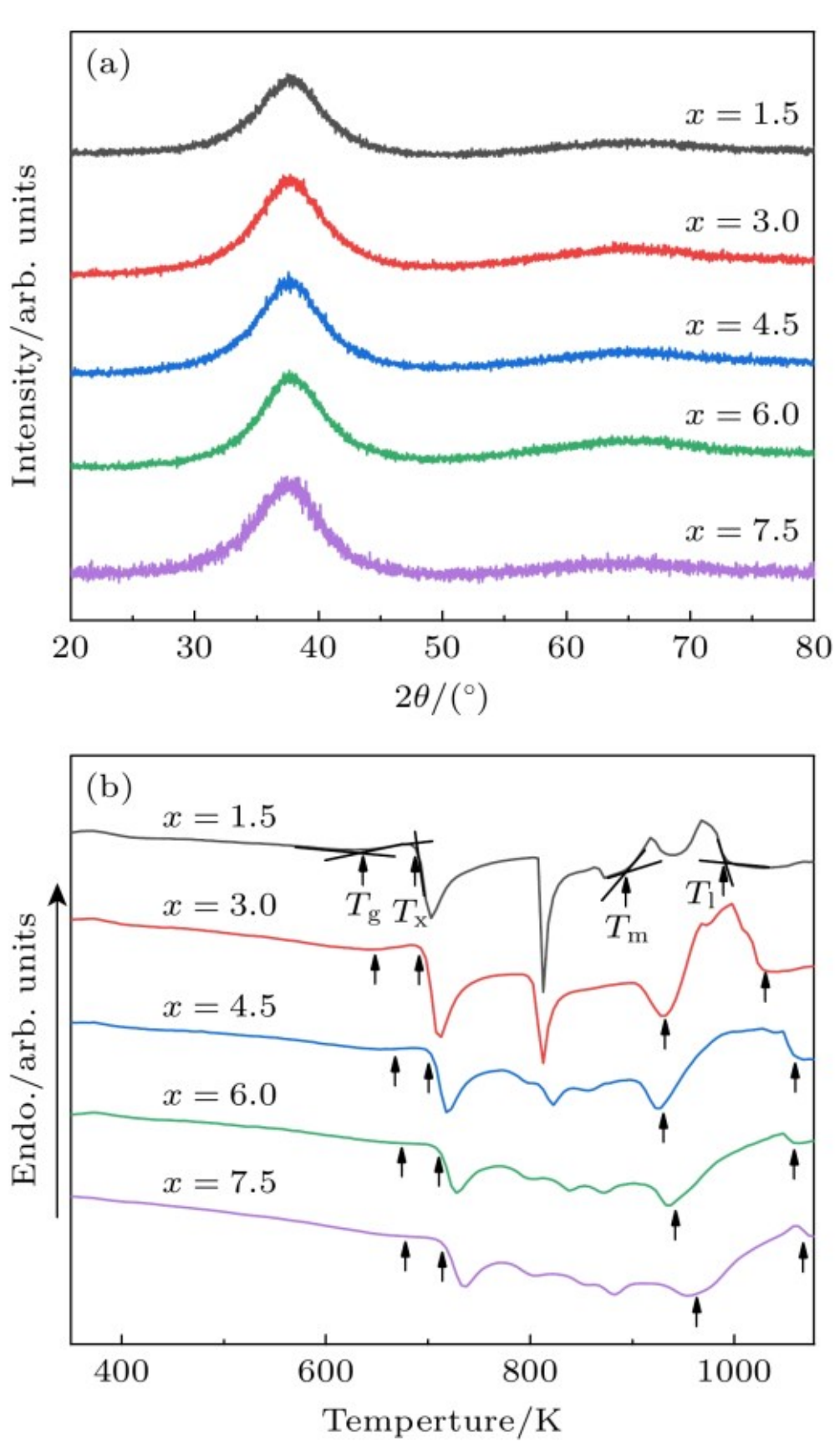

**Fig. 1** (a) XRD spectrum and (b) DSC curve of $((Ti_{40}Zr_{40}Ni_{20})_{72}Be_{28})_{100-x}Al_x$ BMGs with a diameter of 2 mm

**Table 1** Thermal physical parameters of $((Ti_{40}Zr_{40}Ni_{20})_{72}Be_{28})_{100-x}Al_x$ BMGs

| Composition | $T_g$/K | $T_x$/K | $T_m$/K | $T_l$/K | $\Delta T_x$/K | $T_{rg}$ | $\gamma$ |
|---|---|---|---|---|---|---|---|
| $x = 1.5$ | 633.1 | 690.6 | 898.9 | 988.9 | 57.5 | 0.64 | 0.43 |
| $x = 3.0$ | 641.1 | 698.2 | 939.2 | 1026.0 | 57.1 | 0.62 | 0.42 |
| $x = 4.5$ | 644.3 | 705.6 | 930.3 | 1056 | 61.3 | 0.61 | 0.41 |
| $x = 6.0$ | 649.4 | 712.1 | 938.8 | 1089.5 | 62.7 | 0.60 | 0.41 |
| $x = 7.5$ | 655.4 | 718.2 | 976.1 | 1125.3 | 62.8 | 0.58 | 0.40 |

Based on the thermodynamics and kinetics of glass formation, numerous criteria have been proposed to evaluate glass-forming ability (GFA). Among these, the two most commonly used parameters are the reduced glass transition temperature $T_{rg} = T_g/T_l$ and the $\gamma$ parameter defined as $\gamma = T_x/(T_g + T_l)$[15]. The corresponding calculated values are listed in Table 1. As the atomic content of Al increases, the reduced glass transition temperature $T_{rg}$ and the $\gamma$ parameter of the alloys remain within the ranges of 0.58-0.64 and 0.40-0.43, respectively. These relatively high values indicate that the entire alloy system exhibits excellent glass-forming ability. However, the slight decreasing trends in both $T_{rg}$ and $\gamma$ with increasing Al content suggest that the glass-forming ability may marginally decline as the Al concentration rises.

Figure 2 presents the uniaxial compressive stress-strain curves at room temperature for 2 mm diameter $((Ti_{40}Zr_{40}Ni_{20})_{72}Be_{28})_{100-x}Al_x$ BMGs. All alloys yield after reaching their elastic limit and undergo a certain degree of plastic deformation. Following yielding, the alloys exhibit a stage resembling work hardening, where stress gradually increases to a maximum value before softening occurs until fracture. Table 2 summarizes the yield strength ($\sigma_y$), maximum compressive strength ($\sigma_{max}$), and plastic strain ($\varepsilon_p$) for the different alloys. Additionally, alloy densities ($\rho$) were measured using the Archimedes displacement method, and specific strengths ($\sigma_c$) were subsequently calculated; these values are also provided in Table 2. The addition of Al significantly alters the deformation behavior of the alloys. At an Al atomic content of 3%, both the maximum compressive strength and plastic strain reach their peak values of 2845 MPa and 13%, respectively, indicating optimal overall mechanical performance for this composition. With further increases in Al content, both strength and plasticity exhibit a declining trend. Overall, the incorporation of Al induces a characteristic "increase-then-decrease" variation in compressive strength and plastic deformation within the $((Ti_{40}Zr_{40}Ni_{20})_{72}Be_{28})_{100-x}Al_x$ system, with the Al = 3 at.% composition achieving the best balance between strength and ductility.

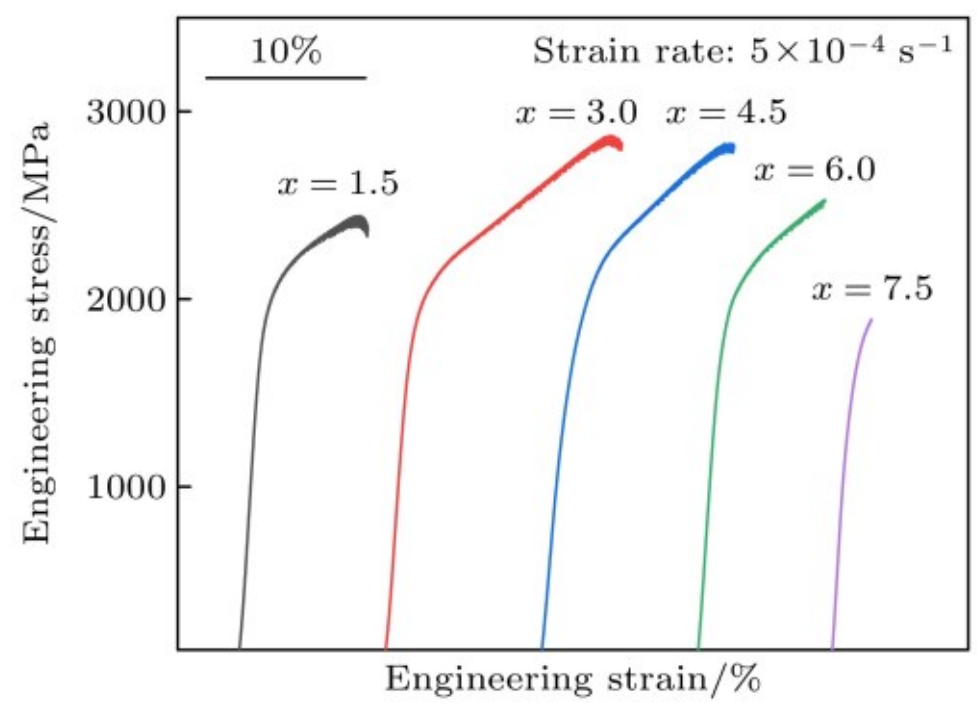

**Fig. 2** The uniaxial compression stress-strain curve of $((Ti_{40}Zr_{40}Ni_{20})_{72}Be_{28})_{100-x}Al_x$ bulk metallic glasses (BMGs) with a diameter of 2 mm

**Table 2** The mechanical properties and density of the $((Ti_{40}Zr_{40}Ni_{20})_{72}Be_{28})_{100-x}Al_x$ bulk metallic glasses (BMGs)

| Alloy | $\sigma_y$/MPa | $\sigma_{max}$/MPa | $\varepsilon_p$/% | $\rho$/(g·cm$^{-3}$) | $\sigma_c$/(N·m·kg$^{-1}$) |
|---|---|---|---|---|---|
| $x = 1.5$ | 1830 ± 51 | 2433 ± 28 | 6.3 ± 1.1 | 5.42 ± 0.1 | $4.49 \times 10^5$ |
| $x = 3$ | 1833 ± 37 | 2845 ± 30 | 13 ± 1.7 | 5.33 ± 0.07 | $5.34 \times 10^5$ |
| $x = 4.5$ | 1822 ± 43 | 2803 ± 32 | 9.6 ± 2.2 | 5.26 ± 0.09 | $5.33 \times 10^5$ |
| $x = 6$ | 1756 ± 21 | 2529 ± 19 | 6.4 ± 1.4 | 5.20 ± 0.1 | $4.86 \times 10^5$ |
| $x = 7.5$ | 1884 ± 28 | 1884 ± 23 | — | 5.16 ± 0.05 | $3.65 \times 10^5$ |

Owing to the fact that plastic deformation in BMGs is almost entirely confined to narrow shear bands, variations in their plasticity are closely linked to shear band evolution. To further investigate the influence of Al addition on the deformation mechanisms of $((Ti_{40}Zr_{40}Ni_{20})_{72}Be_{28})_{100-x}Al_x$ BMGs, we conducted scanning electron microscopy (SEM) observations of the compressive fracture surfaces and lateral shear band morphologies of specimens with varying Al contents, as shown in Figure 3. As illustrated in Figure 3($a_1$), the compressive fracture angle of the Al1.5 specimen is 42.2°, consistent with the von Mises fracture criterion. However, with increasing Al content, the compressive fracture angles of Al3 and Al6 specimens increase to 50.1° and 48.9°, respectively (Figures 3($b_1$) and ($c_1$)), suggesting that the alloys likely experienced significant normal tensile stress during fracture. This tensile stress arises from lateral expansion of the material, indicating that Al addition alters the fracture mechanism[16]. Lateral observations of the specimens reveal that primary shear bands extending along the fracture direction, as well as numerous secondary shear bands formed by branching and intersection, are present in Al1.5 (Figure 3($a_2$)), Al3 (Figure 3($b_2$)), and Al6 (Figure 3($c_2$)). The interaction among multiple shear bands not only impedes rapid shear band propagation but also effectively alleviates localized stress concentration and suppresses crack initiation, thereby substantially enhancing the plastic deformability of BMGs[17]. Among these, the Al3 specimen exhibits the densest shear band network, correlating with its superior plastic performance.

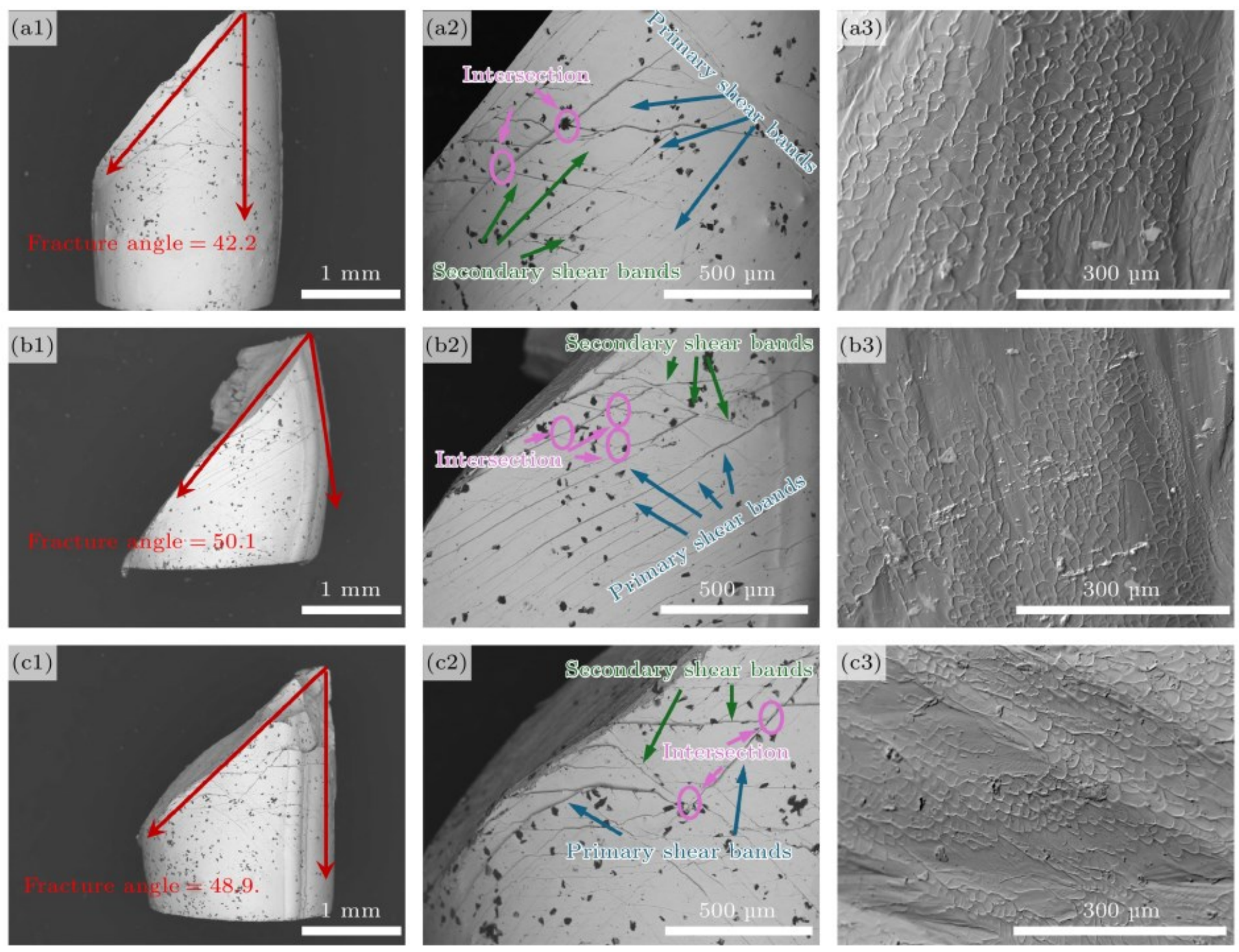

**Fig. 3** The fracture morphology and side shear band morphology of the $((Ti_{40}Zr_{40}Ni_{20})_{72}Be_{28})_{100-x}Al_x$ bulk metallic glasses (BMGs) with a diameter of 2 mm: ($a_1$)-($a_3$) $x = 1.5$; ($b_1$)-($b_3$) $x = 3$; ($c_1$)-($c_3$) $x = 6$

Further observation of fracture morphologies (Fig. 3($a_3$), ($b_3$), ($c_3$)) reveals typical vein-like patterns on all samples. These patterns form due to the instantaneous release of high elastic energy and associated localized adiabatic heating prior to shear band fracture, causing localized softening or melting of the amorphous structure. Spaepen[18] attributes these vein patterns to viscous linkage fracture initiated by shallow voids within shear bands, whereas Argon[19] proposes they result from flow solidification in locally softened regions. Studies indicate that such vein-like patterns can retard catastrophic shear band propagation, thereby enhancing the plasticity of BMGs[20]. Statistical analysis of the average vein pattern dimensions shows values of 18.6, 14.0, and 16.7 μm for the Al1.5, Al3, and Al6 samples, respectively. A clear inverse correlation exists between vein size and alloy plasticity, with the Al3 sample-exhibiting the best plasticity-possessing the smallest vein pattern dimensions.

To further elucidate the comprehensive performance advantages of $((Ti_{40}Zr_{40}Ni_{20})_{72}Be_{28})_{100-x}Al_x$ BMGs relative to conventional metals and other BMGs, specific strength-density and specific strength–plasticity relationships for various materials are plotted in Fig. 4. Figure 4(a) presents the specific strength versus density for Mg-based, Al-based, Ti-based, and Pd-based BMGs, along with the representative Ti-6Al-4V alloy, AZ91 Mg alloy, and 7075-T6 Al alloy[10,21]. The data demonstrate that the $((Ti_{40}Zr_{40}Ni_{20})_{72}Be_{28})_{100-x}Al_x$ BMGs developed in this study exhibit exceptionally high specific strength, surpassing existing records for Ti-based BMGs and significantly exceeding those of conventional Ti, Al, and Mg alloys. Moreover, their relatively low density further underscores their potential as lightweight structural materials. Figure 4(b) illustrates the specific strength versus plasticity for Ti-based, Mg-based, Al-based, and Pd-based BMGs. The results indicate that this alloy system not only achieves ultrahigh specific strength but also demonstrates excellent plasticity, thereby overcoming the traditional trade-off between "high specific strength and low plasticity" in BMGs and successfully balancing both properties.

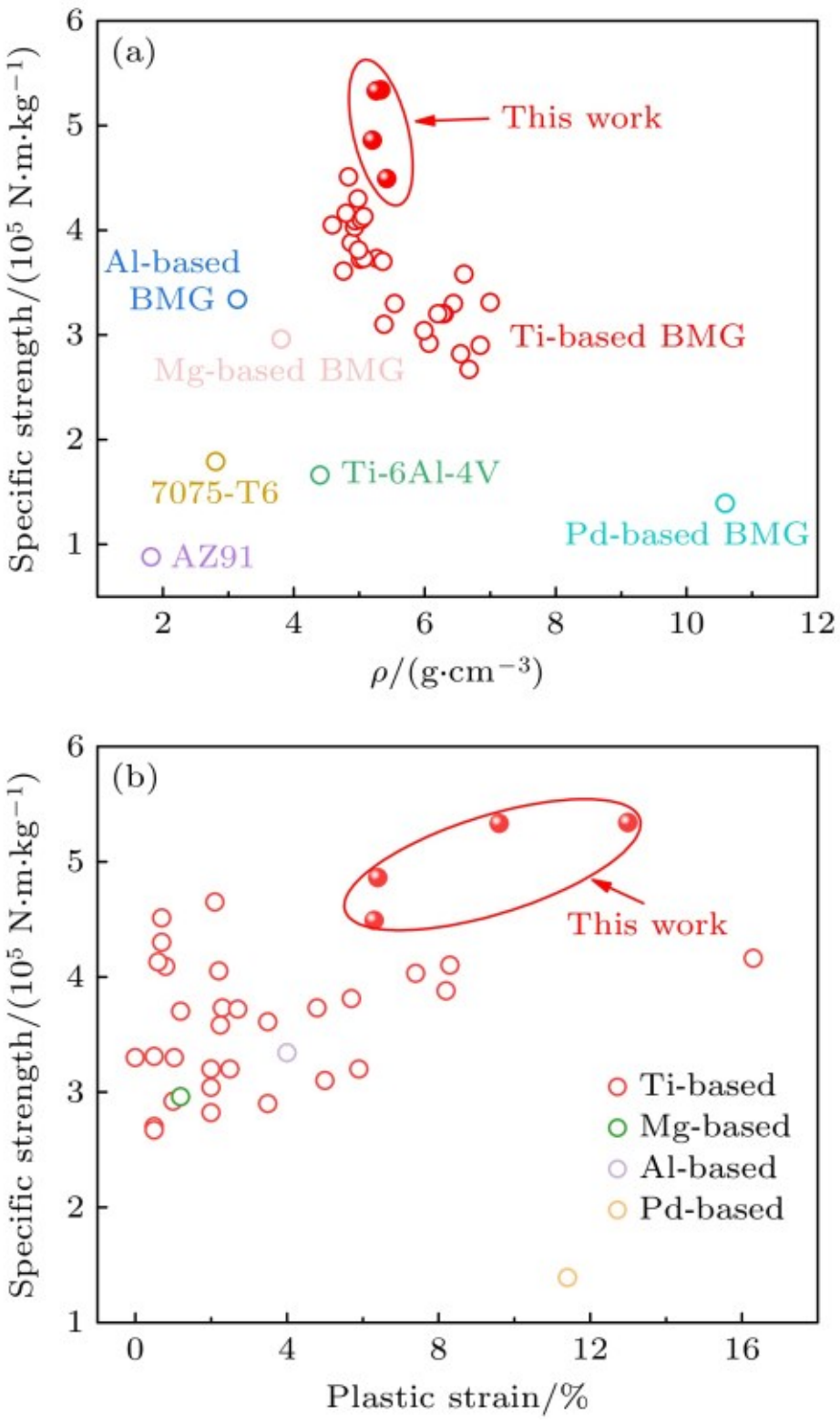

**Fig. 4** Comparison of the properties of the $((Ti_{40}Zr_{40}Ni_{20})_{72}Be_{28})_{100-x}Al_x$ bulk metallic glasses (BMGs) with traditional crystalline alloys and other BMGs: (a) Comparison of density and specific strength with other systems; (b) comparison of plastic deformation ability and specific strength with other systems

## 4 Discussion

The addition of Al significantly enhances the thermodynamic properties and mechanical performance of $((Ti_{40}Zr_{40}Ni_{20})_{72}Be_{28})_{100-x}Al_x$ BMGs. Regarding improved thermal stability, Al incorporation markedly strengthens the resistance of BMGs to crystallization. This aligns with the established mechanism in Zr-based systems, wherein Al promotes short-range ordered structures, thereby enhancing thermal stability[22]. This system employs a $Ti_{40}Zr_{40}Ni_{20}$ quasicrystal as the precursor, inherently rich in icosahedral-like clusters. The addition of Be facilitates amorphization through the "confusion principle" while preserving partially distorted or intact cluster structures[11,23]. Upon further Al addition, the atomic size mismatches of approximately 11.9% and 12.6% between Al and Zr (0.160 nm) and Ni (0.125 nm), respectively[22], promote the formation of densely packed, randomly arranged icosahedral-like clusters. The high density of these clusters substantially increases atomic diffusion resistance, thereby suppressing nucleation and growth of crystalline phases. Concurrently, the strongly negative mixing enthalpies of Al with Zr (-44 kJ/mol), Ti (-30 kJ/mol), and Ni (-22 kJ/mol) favor stable Al-Zr and Al-Ti atomic pairs. These strong bonds act as cluster cores, enhancing the cohesion of icosahedral-like clusters[14] and thus inhibiting long-range ordering while improving overall thermal stability.

In terms of mechanical properties, Al addition achieves simultaneous improvements in specific strength and ductility through the synergistic effects of structural heterogeneity modulation and bond strengthening. First, hybridization between Al 3p orbitals and d orbitals of transition metals such as Zr and Ti forms covalent-like bonds, enhancing structural stability at the electronic level. This is consistent with the reduced Knight shift observed via nuclear magnetic resonance in Al-based systems[23]. Second, the negative mixing enthalpy drives the formation of strong Al-Zr and Al-Ti bonds, constructing a "rigid skeleton" that significantly enhances load-bearing capacity while maintaining low density[24]. Additionally, Al introduction induces pronounced structural heterogeneity. Due to atomic size differences between Al and the matrix elements, local Al segregation creates Al-rich and Al-depleted regions[12]. These heterogeneous zones provide abundant potential sites for shear transformation zone (STZ) nucleation, promoting the formation of multiple shear bands. Simultaneously, densely packed icosahedral-like clusters act as "hard regions" that impede rapid shear band propagation, whereas loosely packed, defect-rich regions accommodate shear strain, causing shear band branching and deflection[14,25-29]. The synergy between these mechanisms significantly delays localized instability, enabling progressive strain transfer and dispersion, and thereby exhibiting work-hardening-like behavior. This work-hardening mechanism is rare in conventional BMGs and endows the alloy with exceptional specific strength and ductility.

## 5 Conclusions

In summary, in $((Ti_{40}Zr_{40}Ni_{20})_{72}Be_{28})_{100-x}Al_x$ BMGs, the quasicrystal precursor preserves numerous distorted or intact clusters. The introduction of Al enables a synergistic interplay among atomic size-mismatch-induced structural heterogeneity, negative mixing enthalpy-driven bond strengthening, and electronic hybridization-induced structural stabilization. These mechanisms collectively regulate the nucleation and evolution of shear bands, ultimately achieving concurrent enhancement of thermal stability and mechanical performance. This provides a novel structural design strategy for lightweight BMGs with high specific strength and ductility. Furthermore, the proposed "quasicrystal-as-precursor" structural modulation approach exhibits broad applicability to other Zr-based and Co-based amorphous alloy systems. Future work should combine experimental measurements with simulation calculations to systematically investigate the evolution of the strain rate sensitivity exponent ($m$). An increase in the $m$ value is expected to enable controllable shear band behavior, facilitating hierarchical strain transfer from localized to global scales, thereby revealing the intrinsic mechanisms underlying ductility enhancement in Al-microalloyed systems. Concurrently, advanced characterization techniques such as atom probe tomography and synchrotron radiation scattering could be employed to further elucidate Al segregation characteristics and their influence on short-range order evolution, thereby deepening the comprehensive understanding of microalloying regulation mechanisms.

1